\documentclass{emulateapj}
\usepackage{graphicx}
\begin{document}

\def\ba{\begin{eqnarray}}
\def\ea{\end{eqnarray}}
\def\etal{et al.\ \rm}

\title{Convective cooling and fragmentation of gravitationally unstable disks.}

\author{Roman R. Rafikov\altaffilmark{1,2}}
\altaffiltext{1}{CITA, McLennan Physics Labs, 60 St. George St., 
University of Toronto, Toronto ON M5S 3H8 Canada; rrr@cita.utoronto.ca}
\altaffiltext{2}{Canada Research Chair}

%%%%%%%%%%%%%%%%%%%%%%%%%%%%%%%%%%%%%%%%%%%%%%%%%%%%%%%%%%%

\begin{abstract}
Gravitationally unstable disks can fragment and form bound
objects provided that their cooling time is short. In 
protoplanetary disks radiative cooling is likely to be 
too slow to permit formation of planets by fragmentation
within several tens of AU from the star. Recently, convection 
has been suggested as a faster means of heat loss from the disk 
but here we demonstrate that it is only marginally more efficient 
than the radiative cooling. The crucial factor is 
the rate at which energy can be radiated from the disk 
photosphere, which is robustly limited from above in the 
convective case by the adiabatic temperature gradient (given a 
certain midplane temperature). Thus, although vigorous 
convection is definitely possible in disks, the inefficiency 
of radiative loss from the photosphere may create a 
bottleneck limiting the ability of the disk to form 
self-gravitating objects. Based on this argument we derive 
a set of analytical constraints which diagnose the 
susceptibility of an unstable disk to fragmentation and 
show that the formation of giant planets by fragmentation 
of protoplanetary disks is unlikely to occur at distances 
of tens of AU. At the same time these constraints do not 
preclude the possibility of fragmentation and star formation 
in accretion disks around supermassive black holes.
\end{abstract}

%%%%%%%%%%%%%%%%%%%%%%%%%%%%%%%%%%%%%%%%%%%%%%%%%%%%%%%%%%%
%%%%%%%%%%%%%%%%%%%%%%%%%%%%%%%%%%%%%%%%%%%%%%%%%%%%%%%%%%%

\section{Introduction.}
\label{sect:intro}

%%%%%%%%%%%%%%%%%%%%%%%%%%%%%%%%%%%%%%%%%%%%%%%%%%%%%%%%%%%

It has been known for a long time (Safronov 1960; Toomre 1964) that 
massive differentially rotating disks are prone to 
gravitational instability (hereafter GI) which sets in
when the destabilizing effect of the self-gravity in the disk 
exceeds the combined restoring action of the pressure and Coriolis 
forces. Instability condition is conveniently expressed in terms of
the so-called Toomre Q-parameter as
\ba
Q\equiv\frac{\Omega c_s}{\pi G\Sigma}\lesssim Q_0,
\label{eq:Toomre}
\ea
where $\Omega,~c_s$, and $\Sigma$ are the local angular frequency 
(disk is assumed to be in Keplerian rotation), sound speed, and 
surface density of the disk, and $Q_0\approx 0.75-1.5$ 
(Kim, Ostriker, \& Stone 2002; Boss 2002) 
is a threshold value of $Q$ at which GI sets in.

Nonlinear outcome of the GI
depends on the thermodynamics of the disk, namely on its
ability to cool rapidly (Gammie 2001; Johnson 
\& Gammie 2003). Rapid cooling keeps pressure forces subdominant 
compared to the disk self-gravity in the non-linear stage of GI 
thus promoting the collapse of unstable parts of the disk and 
leading to the formation of bound objects. When the cooling is 
slow fragmentation is avoided and the disk settles to a quasi-steady 
regime of gravitoturbulence (Gammie 2001) characterized by 
dramatic perturbations of the disk surface density and 
fluid velocity (Durisen \etal 2006). Gammie (2001), Rice \etal
(2003), and Rice, Lodato, \& Armitage (2005) have demonstrated 
using simulations with fixed cooling time $t_{cool}$ that 
fragmentation requires 
\ba
\Omega t_{cool}\lesssim \zeta,
\label{eq:frag_crit}
\ea
where $\zeta\approx 3-12$ depending on the adiabatic index $\gamma$ 
of the disk material, although in some situations $\zeta$ can be 
much higher (Johnson \& Gammie 2003). 

Fragmentation resulting from GI is likely an important 
ingredient of the dynamics of accretion disks
in the central regions of galaxies, around supermassive black 
holes (SMBHs; Paczynski 1978; Kolykhalov \& Sunyaev 1980; 
Illarionov \& Romanova 1988; Levin 2003; Goodman 2003;Goodman \& Tan 2004; 
Thompson, Quataert, \& Murray 2005), including the past
disk around the black hole in the center of our Galaxy 
(Levin \& Beloborodov 2003; Nayakshin 2005). 
On smaller scales, fragmentation of gravitationally unstable  
protoplanetary disks resulting in direct formation of gas giant 
planets has been put forward (Cameron 1978; Boss 1998) as an 
alternative to the conventional core instability model (Perri \& 
Cameron 1974; Harris 1978; Mizuno 1980). Some simulations 
incorporating radiation transfer (Boss 2002; Mayer 2006) 
seem to demonstrate that planet 
formation by GI is possible. However, these results are 
in drastic conflict with the long radiative cooling times 
of optically thick protoplanetary disks (Rafikov 2005; 
hereafter R05) and the recent numerical results by another 
group (Cai \etal 2006; Boley \etal 2006).

To reconcile inefficient radiative cooling with fragmentation 
exhibited in their simulations Boss (2004) and Mayer \etal 
(2006) have come forward with a suggestion that gravitationally unstable 
protoplanetary disks can be efficiently cooled by convection 
so that fragmentation becomes possible. The goal of this paper 
is to critically revise this possibility by setting in 
\S \ref{sect:up_limit} a lower limit on the cooling 
timescale of the disk in which energy is transported away from 
the midplane by convective motions. We then use this 
result in \S \ref{sect:fragmentation} to constrain 
the properties of the disks in which GI and convective 
cooling lead to fragmentation.

%%%%%%%%%%%%%%%%%%%%%%%%%%%%%%%%%%%%%%%%%%%%%%%%%%%%%%%%%%%
%%%%%%%%%%%%%%%%%%%%%%%%%%%%%%%%%%%%%%%%%%%%%%%%%%%%%%%%%%%

\section{Conditions for convection.}
\label{sect:conv}

%%%%%%%%%%%%%%%%%%%%%%%%%%%%%%%%%%%%%%%%%%%%%%%%%%%%%%%%%%%

The question of whether energy in the disk is transported by 
convection or radiation depends on the behavior of disk 
opacity. Bell \& Lin (1994) have demonstrated
that in some temperature intervals opacity $\kappa(P,T)$ can 
be well represented as
\ba
\kappa=\kappa_0 P^\alpha T^\beta,
\label{eq:opacity}
\ea
($P$ and $T$ are the gas pressure and temperature) which we 
assume to hold throughout the paper. 

In this section we focus on optically thick disks, for 
which $\tau\equiv \int_{-\infty}^\infty \rho\kappa dz\gg 1$.
Following the 
approach of Lin \& Papaloizou (1980) we demonstrate in Appendix 
\ref{ap:conditions} that some optically thick disks are 
guaranteed to be convective whenever
\ba
\nabla_0\ge\nabla_{ad}.
\label{eq:conv_inst}
\ea
Here
\ba
\nabla_0\equiv\frac{1+\alpha}{4-\beta},~~~\nabla_{ad}\equiv
\frac{\gamma-1}{\gamma}.
\label{eq:nablas}
\ea
Condition (\ref{eq:conv_inst}) directly follows from 
some assumptions about the spatial distribution of 
the energy dissipation rate within the disk (see Appendix 
\ref{ap:conditions}) and the Schwarzschild criterion 
for convection
\ba
\nabla>\nabla_{ad},~~~\nabla\equiv\frac{d\ln T}{d\ln P},
\label{eq:Sch}
\ea
where $\nabla$ is a temperature gradient. 

Using condition (\ref{eq:conv_inst}) one can infer that in the 
low-temperature regime disk is convective for $T\lesssim 150$ 
K, since then $\kappa$ is dominated by ice grains 
($\alpha=0,~\beta\approx 2$; Bell \& Lin 1994) 
and $\nabla_{ad}=2/7$ while $\nabla_0\approx 1/2$ 
($\gamma=7/5$) in this case. At higher 
temperatures, when ice grains sublimate (but below the sublimation
point of the silicate and metal grains) $\alpha=0,~\beta\approx 1/2$ 
and $\nabla_0\approx \nabla_{ad}$, so that the disk is marginally
unstable for convection. At the same time, submillimeter observations 
of protoplanetary disks demonstrate that dust opacity at
low temperatures is best described by $\alpha=0,\beta\approx 1$ 
(Kitamura \etal 2002), most likely as a result of dust grain growth. 
In this case $\nabla_0=1/3$ and disk is convective.

At higher temperatures, when all dust grains sublimate, 
opacity is pressure-dependent ($\alpha\neq 0$). In particular,
$\alpha\approx 2/3$ and $\beta\approx 7/3$ in the regime of molecular
opacity ($T\approx 1.5-5\times 10^3$ K), so that 
$\nabla_0\approx 1$ and disk is again convective. When using 
criterion (\ref{eq:conv_inst}) at even higher temperatures one has to 
bear in mind possible importance of the radiation pressure 
which we neglect in our analysis.

%%%%%%%%%%%%%%%%%%%%%%%%%%%%%%%%%%%%%%%%%%%%%%%%%%%%%%%%%%%
%%%%%%%%%%%%%%%%%%%%%%%%%%%%%%%%%%%%%%%%%%%%%%%%%%%%%%%%%%%

\section{Upper limit on the convective cooling rate.}
\label{sect:up_limit}

%%%%%%%%%%%%%%%%%%%%%%%%%%%%%%%%%%%%%%%%%%%%%%%%%%%%%%%%%%%

We first consider optically thick disks in which dense midplane 
regions cool in a two-stage fashion. Their heat first needs 
to be transported somehow to the disk photosphere 
(located at $\tau\sim 1$) and then lost to space 
by radiation from the photosphere, irrespective of whether 
transport of energy inside the disk is done by radiation or 
convection. 

In a convective disk the first leg of the cooling 
process can be rather fast but not instantaneous. Indeed, 
convection is driven by the buoyancy of the hot gas, but 
buoyancy cannot supply gas with the vertical acceleration 
higher than vertical gravitational acceleration $g_z$. 
Because of that, when the parcel of hot gas reaches about a  
scale height above the midplane (which is roughly the 
location of the photosphere) its vertical speed would be 
at most $\sim c_s$ and it would take time at least 
$\sim\Omega^{-1}$ to get there. Thus, the heat transport 
inside the disk cannot occur in less than the dynamical 
timescale of the disk. If this were the end of story, 
criterion (\ref{eq:frag_crit}) would suggest that convective 
energy transport is marginally compatible with fragmentation.  

However, the need to eventually lose the transported energy 
by radiation from the photosphere changes this 
conclusion, since the cooling time is in fact the sum of the
interior heat transport timescale (which can be as short as 
$\Omega^{-1}$) and the timescale of radiative losses from 
the disk surface. For a given midplane temperature $T_m$ 
the fastest cooling occurs for the highest possible photospheric 
temperature $T_{ph}$, since the energy loss from the photosphere  
per unit area is $\sigma T_{ph}^4$, assuming blackbody emission.
For a fixed $T_m$ higher $T_{ph}$ implies smaller temperature 
variation for a fixed change in pressure, or, in other words, 
lower value of $\nabla$ [see equation (\ref{eq:Sch})]. 
Thus, disk cools most efficiently when $\nabla$ assumes 
its lowest possible value.

Whenever the radiative temperature gradient $\nabla_r$ defined as
\ba
\nabla_r\equiv\frac{3}{16}\frac{\kappa P F}{g_z \sigma T^4}
\label{eq:rad_grad}
\ea
(here $\sigma$ is a Stephan-Boltzmann constant and 
$F$ is the upward energy flux) satisfies the condition 
$\nabla_r<\nabla_{ad}$, disk is 
convectively stable and $\nabla=\nabla_r<\nabla_{ad}$. In
this case energy is transported from the midplane of the disk
to its photosphere radiatively, and equations (\ref{eq:Toomre})
and (\ref{eq:frag_crit}) directly yield constraints on the
properties of disks capable of forming bound objects
by the GI (R05). As $\nabla_r$ becomes 
larger than $\nabla_{ad}$ convection sets in and the temperature 
gradient obeys $\nabla_{ad}<\nabla<\nabla_r$. From this
one can immediately see that the smallest possible value of
$\nabla$ (leading to highest $T_{ph}$) in a 
convective disk is $\nabla_{ad}$. Situation when 
$\nabla\approx\nabla_{ad}$ is possible only in the case
of very efficient convection when the radiative flux is small 
compared to the convective flux, which typically 
requires high gas density. Whether this is the case depends 
on the situation at hand but using $\nabla=\nabla_{ad}$ 
allows us to obtain a robust upper limit on the cooling 
rate of the disk.  

Disk with $\nabla=\nabla_{ad}$ is vertically isentropic and 
we assume this to hold all the way to the 
photosphere (this is verified in Appendix 
\ref{ap:cool_isentropic}). Then
\ba
\frac{T_{ph}}{T_m}=\left(\frac{P_{ph}}{P_m}\right)^{\nabla_{ad}},
\label{eq:tem_rat}
\ea
where $P_{ph}$ and $P_m$ are the gas pressure at the photosphere 
and at the midplane, respectively. For the purposes of this
calculation it is 
enough to assume that vertical acceleration $g_z$ is roughly 
constant in which case
\ba
\frac{P_{ph}}{P_m}\approx \frac{\kappa(P_{m},T_{m})}
{\kappa(P_{ph},T_{ph})}\tau^{-1}. 
\label{eq:P_rat}
\ea
Then one finds from (\ref{eq:opacity}), (\ref{eq:tem_rat}), and
(\ref{eq:P_rat}) that
\ba
\frac{T_{ph}}{T_m}\approx\tau^{-\eta/4},~~~\eta=\frac{4\nabla_{ad}}
{1+\alpha+\beta\nabla_{ad}}.
\label{eq:t_rat_final}
\ea
Thus, the fastest possible rate $L$ of energy loss per unit 
surface area of an optically thick convective disk is 
$L\approx \sigma T_m^4\tau^{-\eta}$. Cassen (1993) used 
similar line of arguments to calculate $T_{ph}$ for the disk with 
constant opacity (which is, in fact, convectively stable). 
Equation (\ref{eq:t_rat_final}) reduces to 
his result $\eta=4(\gamma-1)/\gamma$ if we set $\alpha=\beta=0$.

In the optically thin case ($\tau\lesssim 1$) disk loses 
energy to space by volumetric 
radiative cooling irrespective of whether it is 
convective or radiative. Then, according to the 
Kirchhoff's law,
%\footnote{Line cooling may be more efficient
%than this estimate predicts, but it requires high
%temperatures, which, coupled with the requirement 
%$\tau\lesssim 1$ (implying low $\Sigma$), would likely 
%make disk gravitationally stable.}
$L\approx \sigma T_m^4\tau$. 
In general, for arbitrary $\tau$, one can interpolate 
between the optically thick and thin regimes so that the 
cooling rate per unit surface area is 
\ba
L\approx\frac{\sigma T_m^4}{f_{c}(\tau)},~~~f_{c}(\tau)=
\chi\tau^{\eta}+\frac{\phi}{\tau},
\label{eq:cooling}
\ea
where $\chi,~\phi\sim 1$ are constants. In Appendix 
\ref{ap:cool_isentropic} we present detailed 
calculation of $\chi$ and show that its value does not 
deviate strongly from unity.

Cooling time is a ratio of the total thermal energy content
per unit surface area of the disk $E_{th}\sim \Sigma c_s^2$
[$c_s\equiv(kT_m/\mu)^{1/2}$ is the isothermal sound speed 
corresponding to the midplane temperature $T_m$, the same 
as used in (\ref{eq:Toomre}); here $k$ is a Boltzmann constant, 
$\mu$ is the mean molecular weight]
to the cooling rate $L$. Using (\ref{eq:cooling}) one 
finds the shortest possible cooling time in a convective disk:
\ba
t_{cool} &=& \frac{\Sigma c_s^2}{\sigma T_m^4}f_{c}(\tau)
\label{eq:t_cool}\\
&\approx &
2\times 10^4~\mbox{yrs}~\frac{\Sigma}{10^3\mbox{g cm}^{-2}}
\left(\frac{T_m}{100~\mbox{K}}\right)^{-3}\frac{f_{c}(\tau)}{10^3},
\nonumber
\ea
where the numerical estimate is for the gas of Solar composition
and one has keep in mind that $f_{c}(\tau)\sim 10^3\gg 1$ for the 
adopted fiducial values of $\Sigma$ and $T$.

When the disk is convectively stable energy is transported 
by radiation, however $t_{cool}$ is still given by the 
expression (\ref{eq:t_cool}) but with $f_{c}$ replaced
with $f_r(\tau)\approx\tau+\tau^{-1}$ (R05). In the optically 
thin case cooling is the same for both modes of heat transfer
while in the optically thick regime one finds 
\ba
\frac{f_{c}(\tau)}{f_r(\tau)}\approx \tau^{-\xi},~~~
\xi=\frac{(4-\beta)(\nabla_0-\nabla_{ad})}{1+\alpha+
\beta\nabla_{ad}}.
\label{eq:f_rat}
\ea
Thus, when the condition (\ref{eq:conv_inst}) is satisfied and 
the disk is convective, fastest possible convective cooling is 
more effective than the radiative cooling would be for the same disk 
parameters ($\Sigma, T_m, \Omega$). However, this improvement 
is not very large for two reasons. First, in real disks 
only the condition $\nabla_{ad}<\nabla<\nabla_r$ weaker than 
$\nabla=\nabla_{ad}$, must be fulfilled, so that the real 
cooling rate is somewhere in between the radiative and
the maximum possible convective regimes. Thus, $t_{cool}$ is likely
to be longer than equation (\ref{eq:t_cool}) predicts. Second,
even the shortest cooling time (\ref{eq:t_cool})
is not so different from radiative $t_{cool}$:
for example, $\xi=3/11$ for $\alpha=0, \beta=2$, while 
for $\alpha=0, \beta=1$ (as supported by 
observations of T Tauri disks) one finds $\xi=1/9$. 
Thus, the difference between the radiative
and convective $t_{cool}$ is a very weak function of $\tau$:
for $\tau=10^3$ the ratio of cooling times in two cases are 
$f_c/f_r\approx 7$ and $\approx 2$ correspondingly.

%%%%%%%%%%%%%%%%%%%%%%%%%%%%%%%%%%%%%%%%%%%%%%%%%%%%%%%%%%%
%%%%%%%%%%%%%%%%%%%%%%%%%%%%%%%%%%%%%%%%%%%%%%%%%%%%%%%%%%%

\section{Implications for disk fragmentation.}
\label{sect:fragmentation}

%%%%%%%%%%%%%%%%%%%%%%%%%%%%%%%%%%%%%%%%%%%%%%%%%%%%%%%%%%%

Combination of gravitational instability condition 
(\ref{eq:Toomre}) with the 
cooling requirement (\ref{eq:frag_crit}) allows one to set
a robust lower limit on $\Sigma$ and $T$
of the disk region which can fragment as a result of GI, which
has been previously done in R05 for the case of 
radiative heat transfer. Repeating his arguments for the case of 
maximally efficient convective cooling [i.e. using eq. 
(\ref{eq:t_cool})] 
we find that self-gravitating objects can form only in
disks satisfying the following properties:
\begin{eqnarray}
&& \Sigma\ge \Sigma_{min}= \Sigma_{inf}[f_c(\tau)]^{1/5},
\label{eq:sig_min}\\
&& T_m\ge T_{min}=T_{inf}[f_c(\tau)]^{2/5},
\label{eq:T_min}
\end{eqnarray}
where
\begin{eqnarray}
&& \Sigma_{inf}\equiv\Omega^{7/5}(\pi G Q_0)^{-6/5}
\left[\frac{1}{\zeta\sigma}
\left(\frac{k}{\mu}\right)^4\right]^{1/5}\nonumber\\
&& \approx 6.6 \times 10^5 \mbox{g cm}^{-2}a_{\rm AU}^{-21/10}
\left(Q_0^{6}\tilde \mu^{4}\zeta\right)^{-1/5}
\left(\frac{M_\star}{M_\odot}\right)^{7/10},
\label{eq:sigma_constr}\\
&& T_{inf}\equiv\Omega^{4/5}
\left(\zeta\pi Q_0 G\sigma
\right)^{-2/5}\left(\frac{k}{\mu}\right)^{3/5}\nonumber\\
&& \approx 5800~ \mbox{K}~a_{\rm AU}^{-6/5}\tilde\mu^{-3/5}
\left(Q_0\zeta\right)^{-2/5}
\left(\frac{M_\star}{M_\odot}\right)^{2/5}.
\label{eq:T_constr}
\end{eqnarray}
Here $a_{\rm AU}$ is the semimajor axis $a$ in units of AU,
$\tilde \mu\equiv \mu/m_H$ is the mean molecular weight
relative to the atomic hydrogen mass  $m_H$, and $M_\star$ is 
the mass of the central star ($M_\odot$ is the Solar mass). 
Needless to say, the condition (\ref{eq:T_min}) has to be 
satisfied simultaneously with (\ref{eq:Toomre}) which 
additionally limits $T$ from above.

One can clearly see that the only difference between the
radiative case studied in R05 and convective case
considered here is in the explicit form of the function
$f_c(\tau)$, which changes to $f_r(\tau)$ in the radiative 
case. Bearing in mind that $f_c(\tau)$ is similar
to $f_r(\tau)$ in that it reaches the minimum value 
$f_c\sim 1$
at $\tau\sim 1$ we conclude that the lowest possible 
({\it infimum}) values of $\Sigma$ and $T$ given by 
(\ref{eq:sigma_constr}) and (\ref{eq:T_constr})
are the same as in the radiative case. Thus, all the 
constraints set in R05 on the basis of $\Sigma_{inf}$ and 
$T_{inf}$ persist in the convective case as well. In 
particular, it follows from (\ref{eq:t_rat_final}) and 
(\ref{eq:T_min}) that $T_{ph}\ge T_{inf}[f_c(\tau)]^{3/20}$,
so that the photospheric temperature of fragmenting 
disk $T_{ph}$ is limited from below by 
at least $T_{inf}$.

%%%%%%%%%%%%%%%%%%%%%%%%%%%%%%%%%%%%%%%%%%%%%%%%%%%%%%%%%%%
%%%%%%%%%%%%%%%%%%%%%%%%%%%%%%%%%%%%%%%%%%%%%%%%%%%%%%%%%%%

\section{Application to protoplanetary disks and 
other environments.}
\label{sect:application}

%%%%%%%%%%%%%%%%%%%%%%%%%%%%%%%%%%%%%%%%%%%%%%%%%%%%%%%%%%%

For molecular gas of
solar composition and $\zeta=10,Q_0=1$ one finds $\Sigma_{inf}
\approx 2\times 10^5$ g cm$^{-2}$ and $T_{inf}\approx 1400$ 
K at $1$ AU. These very extreme conditions [$\Sigma$ exceeds 
the corresponding value in the minimum mass Solar nebula 
(Hayashi 1981) by $\sim 10^2$!] are clearly 
incompatible with the observations of T Tauri disks which
precludes the possibility of giant planet formation by GI 
within several AU from the parent star. 
At larger distances, however, the constraints set by  
$\Sigma_{inf}$ and $T_{inf}$ relax and become more
acceptable from the observational point of view. 
At $10$ AU, for example, $\Sigma_{inf}\approx 
1.6\times 10^3$ g cm$^{-2}$ and $T_{inf}\approx 
100$ K which is at (or above) the uppermost end of the
observed distributions of $\Sigma$ and $T$. 

However, in such environment 
$f_c(\tau)\approx\tau\sim 10^3\gg 1$ 
(for $\kappa\approx 1$ cm$^2$ g$^{-1}$) so that equations 
(\ref{eq:sig_min})-(\ref{eq:T_min}) set much more severe 
constraints on the disk properties. Careful inspection of 
(\ref{eq:sig_min})-(\ref{eq:T_min}) demonstrates that 
planet formation by GI at $10$ AU is only possible 
if $T>1500$ K so that dust has sublimated lowering 
the opacity --- otherwise $\tau$ would be so high that 
according to (\ref{eq:T_min}) $T$ cannot 
self-consistently remain below the sublimation 
point. Thus, planet formation by GI even at $10$ AU from 
the central star requires disk properties 
which are clearly inconsistent with the current 
observations of protoplanetary disks 
(Kitamura \etal 2002). Only beyond $20$ AU do we find
that $T_{min}<1500$ K and $\Sigma_{min}<2.4\times 10^3$ 
g cm$^{-2}$ so that silicates condense. At $45$ AU 
$T\lesssim 150$ K and ice grains condense which 
increases opacity by a factor of several. In general, 
we find that the disk properties limited by 
(\ref{eq:sig_min})-(\ref{eq:T_min}) become more 
or less compatible with the observations of
T Tauri disks only at $a\sim 100$ AU. Only this far 
from the star can planets possibly form by the 
GI, which contradicts the conclusion of Boss (2006).

Previous discussion focused on the local disk
properties necessary for planet formation by GI. It is 
however clear, by analogy with R05, that the global
parameters of the fragmenting disk -- its mass and 
luminosity -- should also be rather extreme. We find 
that the disk mass would exceed several $M_\odot$ if planets were 
to form by GI at $20$ AU, but it reduces to 
$\sim 0.1~M_\odot$ at $100$ AU. The masses of 
self-gravitating objects that would form as a result
of fragmentation are also going to be large, likely
in the brown dwarf regime (R05), although this can
be reliably clarified only by numerical 
simulations. 

Fragmentation of gravitationally unstable disks around 
SMBHs in galactic centers is also constrained by the 
conditions (\ref{eq:sig_min})-(\ref{eq:T_min}) if 
these disks are convective. In this case, unlike the situation 
with the protoplanetary disks, no obvious conflict 
with the observations arises. Indeed, let's 
consider the possibility of star formation in a disk at
$a=0.1$ ps from the SMBH in 
the Galactic Center. For a black hole mass of 
$3\times 10^6~M_\odot$ simultaneous fragmentation and 
GI at $0.1$ pc require $\Sigma_{inf}\approx 6$ g cm$^{-2}$
and $T_{inf}\approx 4$ K. The real temperature is in 
fact limited from below by the stellar irradiation at the 
level $\sim 40$ K (Nayakshin 2005) and this automatically increases
minimum $\Sigma$ to $\sim 25$ g cm$^{-2}$ to satisfy 
the Toomre criterion (\ref{eq:Toomre}). This translates 
into $\tau\approx 8$ and cooling time $\approx 0.2\Omega^{-1}$ 
which would lead to fragmentation. 
Characteristic disk mass $\pi a^2\Sigma \sim 4\times 
10^3~M_\odot$ is fully compatible with the 
current constraints on the total stellar mass in the
two stellar disks around the central black hole in
the Galactic Center (Nayakshin \etal 2006).

%%%%%%%%%%%%%%%%%%%%%%%%%%%%%%%%%%%%%%%%%%%%%%%%%%%%%%%%%%%
%%%%%%%%%%%%%%%%%%%%%%%%%%%%%%%%%%%%%%%%%%%%%%%%%%%%%%%%%%%

\section{Discussion.}
\label{sect:disc}

%%%%%%%%%%%%%%%%%%%%%%%%%%%%%%%%%%%%%%%%%%%%%%%%%%%%%%%%%%%

There is good agreement between the constraints 
on $\Sigma$ and $T$ presented in this paper and R05 
and the results of the three-dimensional 
grid-based simulations by Mejia (2004), Cai \etal (2006) 
and Boley \etal (2006). Using radiation transfer scheme 
based on the flux-limited diffusion these authors 
find that planet formation by GI does not occur at
distances of $\sim 10$ AU from the star, in accord with
our findings. This outcome is clearly caused by the long 
radiative cooling timescales in dense and cold disk models 
that are run by these authors. In particular, Cai \etal 
(2006) find that lowering disk metallicity has the effect of
decreasing the cooling timescale, which is obvious from 
equation (\ref{eq:t_cool}) if one also notices that 
the low-temperature opacity $\kappa$ is proportional 
to metallicity and $f_c(\tau)\propto \kappa^{\eta}$ in 
the optically thick regime (typical at distances of 
several tens of AU in massive protoplanetary disks).

At the same time recent radiative hydrodynamical 
simulations by Boss (2004) and Mayer 
\etal (2006) demonstrate efficient fragmentation of 
massive and cold protoplanetary disks and thus 
disagree with both the 
analytical constraints presented in this paper and  
the simulations of Mejia (2004), Cai \etal (2006) 
and Boley \etal (2006). The arguments presented here 
do not support the idea of Boss (2004) and Mayer 
\etal (2006) that energy transport by convection 
can significantly speed up the disk cooling and 
lead to its fragmentation. While there are many factors 
that can cause this discrepancy (opacities, treatment of
shock dissipation, numerical scheme employed, etc.) we
believe that the major source of disagreement is in the 
treatment of the external radiative boundary condition 
by different groups.
 
To illustrate this point we note that Boss (2004), 
Mayer \etal (2006), and Boley \etal (2006) do 
observe\footnote{Boley \etal (2006) find convection 
to be disrupted during the active phase of the GI.} 
efficient convection in their simulations. However, 
rapid vertical transport of energy from the midplane 
to the disk surface is a necessary but not sufficient
condition for rapid cooling, since this energy must
then be radiated away from the photosphere. Thus, the 
photosphere presents a bottleneck for the disk cooling
such that the disk cannot cool faster than its 
photospheric temperature permits. For this reason 
it is very important that the external radiative boundary
condition is treated with extreme 
care\footnote{In particular, the proper treatment of 
the photospheric transition between the optically 
thick and thin regions should be especially challenging
for the SPH simulations such as used by Mayer \etal 
(2006).} in simulations. 

Some of the vertical motions seen in simulations and 
attributed to convection may actually be due to the
upward propagating waves launched by the overdensities 
caused by the GI (Boley \etal 2006). The dissipation of 
such waves may heat 
up the upper layers of the disk which has an effect 
of increasing the photospheric temperature. This, however,
has nothing to do with the cooling of transiently 
collapsing overdensities in the disk as these waves 
do nothing to transport the thermal energy away from 
the dense gas near the midplane. Thus, such wave motions 
cannot facilitate fragmentation of an unstable disk.

Similarly, the external irradiation of the disk (by
the central star or the surrounding envelope) which acts 
to increase the photospheric temperature does not 
accelerate disk cooling. In this case, although the disk 
surface loses more energy because of the higher $T_{ph}$
it also gains energy from the absorbed external radiation
which keeps the net cooling rate unchanged. Cooling of  
overdensities forming near the midplane as a result of GI 
depends only on the temperature gradients that develop 
in the interior and thus is not affected by external
heating. Besides, external illumination  
tends to stabilize temperature gradient and to suppress 
convection.

Our final comment refers to the use of constant factor
$\zeta\sim 1-10$ in the fragmentation condition 
(\ref{eq:frag_crit}). Using simulations with nonlinear 
cooling rates devised to mimic the effect of realistic 
opacities on disk cooling Johnson \& Gammie (2003) have 
demonstrated that $\zeta$ can be significantly larger
than $\sim 10$ just below the so-called opacity gaps ---
regions in the $\rho-T$ space where opacity suddenly 
changes. In particular, they find that $\zeta \sim 10^4$
at $T\sim 10^3$ K, near the point of silicate dust 
sublimation where opacity drops by a factor of 
$\sim 10^3$. Such an increase of $\zeta$ significantly 
relaxes our constraints 
(\ref{eq:sigma_constr})-(\ref{eq:T_constr}) which seems 
to allow fragmentation in disks of lower $\Sigma$ and 
$T$. However, the less conservative (but still robust) 
constraint (\ref{eq:sig_min})-(\ref{eq:T_min}) eliminates 
this concern because (similar to $\zeta$) $\kappa$ and 
$\tau$ are also very large  on the verge of the opacity 
gap (compared to their values at the bottom of the gap) 
so that $f_c\gg 1$ in (\ref{eq:sig_min})-(\ref{eq:T_min}) 
and this offsets the decrease of 
$\Sigma_{inf}$ and $T_{inf}$ driven by large $\zeta$.
Besides, this issue only arises if $T\sim 10^3$ K which 
is atypical for a protoplanetary disk in T Tauri phase 
anyway. 

The arguments presented in this paper and R05 are
analytic in nature, however, they are based solely 
on the numerically verified criteria (\ref{eq:Toomre}) 
and (\ref{eq:frag_crit}). Thus, keeping in mind previous 
discussion, we conclude that to within factors of order 
unity our constraints (\ref{eq:sig_min})-(\ref{eq:T_min}) 
on the properties of unstable disks capable to support 
the formation of self-gravitating objects should be valid 
in applications to realistic disks in which energy is 
transported by convection.

%%%%%%%%%%%%%%%%%%%%%%%%%%%%%%%%%%%%%%%%%%%%%%%%%%%%%%%%%%%
%%%%%%%%%%%%%%%%%%%%%%%%%%%%%%%%%%%%%%%%%%%%%%%%%%%%%%%%%%%

\section{Conclusions.}
\label{sect:concl}

%%%%%%%%%%%%%%%%%%%%%%%%%%%%%%%%%%%%%%%%%%%%%%%%%%%%%%%%%%%

We have investigated the possibility of rapid disk 
cooling by convection in the context of fragmentation of 
gravitationally unstable disks. We have shown that 
even the most extreme form of convective cooling 
(realized when $\nabla=\nabla_{ad}$) produces cooling 
rates which are only marginally higher than in the case 
of radiative energy transport. The major reason for the 
inefficient cooling of dense and cold convective 
disks is in the bottleneck caused by 
the need to eventually radiate from the photosphere the 
energy that has been transported there from the midplane,
irrespective of how the latter has been done. Armed with 
this knowledge we demonstrate that inefficient cooling 
precludes direct formation of giant planets by GI in 
protoplanetary disks anywhere within several tens of
AU from the parent star (analogous conclusion has been 
reached in R05). At the same time, disk cooling 
(convective or radiative) is fast enough to allow 
fragmentation of gravitationally unstable disks around 
black holes in the galactic centers, supporting the 
idea that the efficient star formation is possible 
in such disks.

\acknowledgements 

I am grateful to Chris Matzner and Sergey Nayakshin for 
careful reading of the manuscript and making very useful 
suggestions, to Yuri Levin for helpful exchanges, and to Charles
Gammie for bringing the work by Cassen (1993) to my 
attention. The financial support for this work is provided
by the Canada Research Chairs program and a NSERC 
Discovery grant.

\appendix

%%%%%%%%%%%%%%%%%%%%%%%%%%%%%%%%%%%%%%%%%%%%%%%%%%%%%%%%%%%
%%%%%%%%%%%%%%%%%%%%%%%%%%%%%%%%%%%%%%%%%%%%%%%%%%%%%%%%%%%

\section{Derivation of condition (\ref{eq:conv_inst}).}
\label{ap:conditions}

%%%%%%%%%%%%%%%%%%%%%%%%%%%%%%%%%%%%%%%%%%%%%%%%%%%%%%%%%%%

To derive the condition (\ref{eq:conv_inst}) in the case of opacity 
given by equation (\ref{eq:opacity}) we will follow the 
approach adopted by the Lin \& Papaloizou (1980). Equation of 
hydrostatic equilibrium states that
\ba
\frac{dP}{dz}=-\rho g_z,
\label{eq:hydrostat}
\ea
where vertical acceleration $g_z=\Omega^2 z+g_z^{sg}$ is
contributed both by the central object ($\Omega^2 z$ term) and
the disk self-gravity ($g_z^{sg}$ term). The latter contribution 
satisfies Poisson equation
\ba
\frac{d g_z^{sg}}{d z}=4\pi G\rho.
\label{eq:poisson}
\ea
Radiative transport is described by
\ba
\frac{16\sigma T^3}{3\kappa\rho}\frac{dT}{dz}=-F,~~~~
\frac{dF}{dz}=\epsilon,
\label{eq:rad_tr}
\ea
where $F$ is the vertical radiative flux and $\epsilon$ is the
volumetric energy dissipation rate. Equations (\ref{eq:hydrostat}) 
and (\ref{eq:rad_tr}) result in $\nabla_r$ given by equation 
(\ref{eq:rad_grad}), but they can also be combined in the 
following way: 
\ba
\nabla_0\frac{16}{3}\frac{\sigma}{\kappa_0}
\frac{d(T^{4-\beta})}{d(P^{1+\alpha})}=\frac{F}{g_z}.
\label{eq:rel1}
\ea
We now note that if $F/g_z$ monotonically decreases as 
$z$ increases (and, accordingly, as $P$ decreases), 
then, integrating equation (\ref{eq:rel1}) down from the 
photosphere we arrive at the following inequality:
\ba
\nabla_0\frac{16}{3}\frac{\sigma}{\kappa_0}
\left(T^{4-\beta}-T_{ph}^{4-\beta}\right)=
\int\limits_{P_{ph}}^P\frac{F}{g_z}d(P^{1+\alpha})<
\frac{F}{z}\left(P^{1+\alpha}-P_{ph}^{1+\alpha}\right),
\label{eq:ineq}
\ea  
since $F/g_z$ is assumed to be largest at $P$. This
can be finally rewritten as
\ba
\nabla_0\left[1-\left(\frac{T_{ph}}{T}\right)^{4-\beta}
\right]<\nabla_r\left[1-\left(\frac{P_{ph}}{P}\right)^{1+\alpha}
\right],
\label{eq:nabla_constr}
\ea
which reduces to $\nabla_0<\nabla_r$ at high depth in the 
disk, where $T\gg T_{ph}$ and $P\gg P_{ph}$. 
Thus, if (1) the condition (\ref{eq:conv_inst}) is
fulfilled and (2) $F/g_z$ is a decreasing function of $z$, 
then $\nabla_r>\nabla_{ad}$ and convection has 
to operate in the bulk of the disk since, according to the 
Schwarzschild criterion (\ref{eq:Sch}), disk cannot be 
radiative. 

We now find out under which circumstances $F/g_z$ 
monotonically decreases as $z$ grows. One has
\ba
\frac{g_z^2}{\Omega^2}\frac{d}{dz}\left(\frac{F}{g_z}\right)=
\int\limits_0^z\left[\epsilon(z)-\epsilon(z^\prime)\right]
dz^\prime+\frac{4\pi G}{\Omega^2}\int\limits_0^z
\left[\epsilon(z)\rho(z^\prime)-
\epsilon(z^\prime)\rho(z)\right]dz^\prime,
\label{eq:deriv}
\ea
where equation (\ref{eq:poisson}) has been used. The first term 
in the right-hand side is clearly negative when a reasonable
and weak assumption of $\epsilon$ decreasing with increasing 
$z$ is made. Sign of the second term, which quantifies the 
effect of the disk self-gravity on stability, depends on how rapidly
$\epsilon$ decreases with $z$. If, like in a conventional
$\alpha$-disk (Shakura \& Sunyaev 1973) with constant 
$\alpha$ and sound speed, $\epsilon(z)\propto \rho(z)$, 
then the second term is identically zero and $d(F/g_z)/dz<0$. 
When $\epsilon(z)$ decays with $z$ faster than $\rho(z)$ the
second term in (\ref{eq:deriv}) is negative and 
$d(F/g_z)/dz$ is again negative. However, if 
$\epsilon(z)/\rho(z)$ is an increasing function of $z$ then
the sign of the second term in (\ref{eq:deriv}) is positive
and one cannot tell for sure how $F/g_z$ behaves as a function
of $z$. In the latter case condition (\ref{eq:conv_inst}) may or 
may not be sufficient to determine whether disk is convective.

It may be possible that gravitationally unstable disks
have  $\epsilon(z)/\rho(z)$ increasing with $z$. This can 
be caused, for example, by dissipation of the upward-propagating
waves driven by the instability, which would tend to deposit their
energy high up in the disk stabilizing its structure. We 
can then conclude that in many situations (certainly in 
non-self-gravitating disks) condition (\ref{eq:conv_inst}) is 
sufficient for convection, while in some 
self-gravitating disks a more conservative criterion than 
(\ref{eq:conv_inst}) may be required.

%%%%%%%%%%%%%%%%%%%%%%%%%%%%%%%%%%%%%%%%%%%%%%%%%%%%%%%%%%%
%%%%%%%%%%%%%%%%%%%%%%%%%%%%%%%%%%%%%%%%%%%%%%%%%%%%%%%%%%%

\section{Cooling of an isentropic disk.}
\label{ap:cool_isentropic}

%%%%%%%%%%%%%%%%%%%%%%%%%%%%%%%%%%%%%%%%%%%%%%%%%%%%%%%%%%%

Here we calculate the structure and the cooling time of a
disk that has an isentropic interior (for which the condition 
$\nabla=\nabla_{ad}$ is satisfied) with equation of state 
$P=K\rho^\gamma$ smoothly matching the
outer radiative layer near the photosphere. We will separately 
consider structure of non-self-gravitating disks with 
$g_z=\Omega^2 z$ and self-gravitating disks with $g_z=g_z^{sg}$
satisfying equation (\ref{eq:poisson}).

%%%%%%%%%%%%%%%%%%%%%%%%%%%%%%%%%%%%%%%%%%%%%%%%%%%%%%%%%%%
\subsection{Non-self-gravitating disk.}
\label{ap:nsg_disk}

In the non-self-gravitating isentropic disk equation 
(\ref{eq:hydrostat}) can be solved using $g_z=\Omega^2 z$,
resulting in the following disk structure:
\ba
\frac{T(z)}{T_m}=\left[\frac{P(z)}{P_m}\right]^{\nabla_{ad}}=
\left[\frac{\rho(z)}{\rho_m}\right]^{\gamma-1}=
1-\frac{z^2}{H^2},~~~~~H^2=\frac{2}{\nabla_{ad}}
\frac{k_B T_m}{\mu\Omega^2},
\label{eq:rels}
\ea
where $T_m, P_m, \rho_m$ are the midplane temperature, 
pressure, and density, while $H$ is recognized as the height 
of the disk surface. Using (\ref{eq:rels}) and assuming disk to 
be optically thick (so that the isentropic part contains most of the 
mass) one can calculate the total surface density
\ba
\Sigma=2\int\limits_0^\infty \rho dz=I_1\rho_m H,~~~
I_1\equiv 2\int\limits_0^1 (1-x^2)^{1/(\gamma-1)}dx,
\label{eq:surf_dens}
\ea
and the full optical depth
\ba
\tau=2\int\limits_0^\infty \rho\kappa dz=I_2\rho_m \kappa_m H,~~~
I_2\equiv 2\int\limits_0^1 
(1-x^2)^{\beta+(\alpha\gamma+1)/(\gamma-1)}dx,
\label{eq:opt_depth}
\ea
where $\kappa_m=\kappa(P_m,T_m)=\kappa_0 P_m^\alpha T_m^\beta$ 
is the opacity at the midplane. The total thermal energy content
per unit surface area is
\ba
E_{th}=\frac{2}{\gamma-1}\int\limits_0^\infty P dz=I_3 P_m H,~~~
I_3\equiv \frac{2}{\gamma-1}\int\limits_0^1 
(1-x^2)^{1/\nabla_{ad}}dx,
\label{eq:thermal_energy}
\ea
while the midplane pressure $P_m=(\nabla_{ad}/2)\Omega^2 
H^2\rho_m$ [following from eq. (\ref{eq:rels}) and the ideal 
gas law] can be expressed through the value of $g_z$ at the
disk surface $g_z(H)=\Omega^2H$ in the following way:
\ba
P_m=I_4\rho_m H g_z(H),~~~
I_4\equiv \frac{\nabla_{ad}}{2}.
\label{eq:P_m}
\ea

%%%%%%%%%%%%%%%%%%%%%%%%%%%%%%%%%%%%%%%%%%%%%%%%%%%%%%%%%%%
\subsection{Self-gravitating disk.}
\label{ap:sg_disk}

In a self-gravitating case one can solve the system 
(\ref{eq:hydrostat})-(\ref{eq:poisson})
with the equation of state $P=K\rho^\gamma$ to find
the following implicit relation between $\rho$ and $z$:
\ba
z=H I_0^{-1}\int\limits^1_{\rho/\rho_m}x^{\gamma-2}
\left(1-x^\gamma\right)^{-1/2}dx,
\label{eq:rel_sg}
\ea
where 
\ba
H=\gamma\rho_m^{(\gamma-2)/2}\left(\frac{K}{8\pi G}\right)^{1/2}
I_0,~~~I_0\equiv\int\limits^1_0 x^{\gamma-2}
\left(1-x^\gamma\right)^{-1/2}dx,
\label{eq:height_sg}
\ea
is the height of the disk surface. Using (\ref{eq:rel_sg})
we can derive expressions for $\Sigma, \tau, E_{th}, P_m$
identical to equations (\ref{eq:surf_dens})-(\ref{eq:P_m})
but with constant coefficients $I_k, k=1,..,4$ which are 
now given by 
\ba
&& I_1\equiv 2 I_0^{-1}\int\limits_0^1 
x^{\gamma-1}(1-x^\gamma)^{-1/2}dx,
\label{eq:surf_dens_sg}\\
&& I_2\equiv 2I_0^{-1}\int\limits_0^1 
x^{\alpha\gamma+(\beta+1)(\gamma-1)}(1-x^\gamma)^{-1/2}dx,
\label{eq:opt_depth_sg}\\
&& I_3\equiv \frac{2I_0^{-1}}{\gamma-1}\int\limits_0^1 
x^{2(\gamma-1)}(1-x^\gamma)^{-1/2}dx,
\label{eq:thermal_energy_sg}\\
&& I_4\equiv \frac{4}{\gamma^2 I_0^2 I_1}.
\label{eq:P_m_sg}
\ea
To derive (\ref{eq:P_m_sg}) one should notice that 
$g_z(H)=2\pi G\Sigma$ while $P_m=K\rho_m^\gamma=8\pi G\rho_m^2H^2
(\gamma I_0)^{-2}$, see equation 
(\ref{eq:height_sg}).

%%%%%%%%%%%%%%%%%%%%%%%%%%%%%%%%%%%%%%%%%%%%%%%%%%%%%%%%%%%
\subsection{Outer radiative layer.}
\label{ap:rad_layer}

Somewhere near the disk surface, at $z\approx H$, a transition 
from the convective interior to the radiative atmosphere occurs. 
We assume that this happens at the 
temperature $T_{tr}$ and pressure $P_{tr}$. Inside the 
radiative layer
\ba
T^4=\frac{3}{4}T_{ph}^4\left(\tau+\frac{2}{3}\right).
\label{eq:rad_sol}
\ea
Differentiating this equation with respect to $z$ and using equation
(\ref{eq:hydrostat}) with constant $g_z(H)$ (which is 
justified by the small thickness of the radiative layer compared 
to $H$, following from the assumption $T_{ph}\ll T_m$) one finds that
\ba
T=T_{ph}\left[2^{(\beta-4)/4}+X\right]^{1/(4-\beta)},~~~
X\equiv \frac{3\kappa_0 T_{ph}^\beta}{16\nabla_0 g_z(H)}
P^{1+\alpha}
\label{eq:TofP}
\ea
in the outer radiative zone, so that
\ba
\tau(z)=\int\limits_z^\infty \kappa\rho dz=
\frac{4}{3}\left[\left(2^{(\beta-4)/4}+X\right)^{4/(4-\beta)}
-\frac{1}{2}\right],
\label{eq:tau_z}
\ea
[here we have used equation (\ref{eq:hydrostat})] and
\ba
\nabla_r=\frac{\partial\ln T}{\partial\ln P}=\nabla_0
\frac{X}{2^{(\beta-4)/4}+X}.
\label{eq:nabla_rad}
\ea
Transition from radiative to convective energy transport occurs
when $\nabla_r=\nabla_{ad}$, i.e. at 
\ba
X_{tr}=\frac{2^{(\beta-4)/4}
\nabla_{ad}}{(\nabla_0-\nabla_{ad})},
\label{eq:X_tr} 
\ea
while the photosphere 
($\tau=2/3$) lies at $X_{ph}=1-2^{(\beta-4)/4}$, see equation 
(\ref{eq:tau_z}). 

From equations (\ref{eq:rels}), (\ref{eq:TofP}),
and (\ref{eq:X_tr}) one finds
\ba
&& T_{tr}=\frac{T_{ph}}{2^{1/4}}\left(\frac{\nabla_0}
{\nabla_0-\nabla_{ad}}\right)^{1/(4-\beta)},
\label{eq:T_tr}\\
&& P_{tr}=\left[\frac{16\times 2^{(\beta-4)/4}\nabla_0\nabla_{ad}}
{3(\nabla_0-\nabla_{ad})}\frac{g_z(H)}{\kappa_0 T_{ph}^\beta}
\right]^{1/(1+\alpha)}.
\label{eq:P_tr}
\ea
Using these relations, equations (\ref{eq:opt_depth}) 
and (\ref{eq:P_m}), and the fact that
$T_{tr}/T_m=(P_{tr}/P_m)^{\nabla_{ad}}$ one obtains
\ba
\frac{T_{ph}}{T_m}=\lambda\tau^{-\eta/4},~~~~\lambda=
\left[\left(\frac{16\nabla_{ad} I_2}{3I_4}\right)^{\nabla_{ad}}
\left(\frac{2^{(\beta-4)/4}\nabla_0}{\nabla_0-\nabla_{ad}}
\right)^{\nabla_{ad}-\nabla_0}\right]^{1/(1+\alpha+\beta\nabla_{ad})},
\label{eq:tem_rat_a}
\ea
where $\eta$ was introduced in equation (\ref{eq:t_rat_final}).
%This formula provides accurate relation between the midplane
%temperature of the disk and its photospheric temperature, which
%determines disk cooling.
Using equations (\ref{eq:opt_depth}), (\ref{eq:thermal_energy}) 
and (\ref{eq:tem_rat_a}) we finally derive cooling time in
the optically thick case as
\ba
t_{cool}=\frac{E_{th}}{2\sigma T_{ph}^4}=
\chi\frac{\Sigma c_{s}^2}{\sigma T_m^4}\tau^{\eta},
~~~~~~\chi=\frac{I_3}{2I_1\lambda^4}.
\label{eq:tcool}
\ea
In particular, in the case of a non-self-gravitating disk  
one finds for $\alpha=0,\beta=2$ ($\eta=8/11$) using 
equations (\ref{eq:surf_dens})-(\ref{eq:P_m}) that
$\lambda=1.37, \chi=0.31$, while for $\alpha=0,\beta=1$ 
($\eta=8/9$) one gets $\lambda=1.55, \chi=0.19$. When the disk
is self-gravitating one obtains using equations
(\ref{eq:surf_dens_sg})-(\ref{eq:P_m_sg}) that 
$\lambda=1.24, \chi=0.46$ for $\alpha=0,\beta=2$, while  
$\lambda=1.37, \chi=0.3$ for $\alpha=0,\beta=1$. In realistic
unstable Keplerian disk the self-gravity of the disk and 
the central star contribute roughly equally to $g_z$, 
so that one should expect
$\lambda=1.24-1.37, \chi=0.31-0.46$ for $\alpha=0,\beta=2$
and $\lambda=1.37-1.55, \chi=0.19-0.3$ for $\alpha=0,\beta=1$.
Note that the small value of $\chi$ in some cases should 
not strongly affect our estimates of $\Sigma_{min}$ and 
$T_{min}$ as these quantities depend on $\chi$ rather 
weakly, see equations (\ref{eq:sig_min})-(\ref{eq:T_min}). 

%%%%%%%%%%%%%%%%%%%%%%%%%%%%%%%%%%%%%%%%%%%%%%%%%%%%%%%%%%%

\end{document}